\documentclass[preprint, showkeys]{revtex4}
\usepackage{epsfig}
\usepackage{graphicx}
\usepackage{psfrag}
\usepackage{setspace}
\usepackage{multirow}
\begin{document}

\title{Daily Variations of the Geomagnetic Field in the Brazilian zone.}
\author{D. Oliva$^{1}$, M. A. Esp\'{\i}rito Santo$^{1,2}$ and A. R. R. Papa$^{1,3,\dag}$}
\address{\vskip 0.5cm $^1$ Observat\'orio Nacional,  RJ, 20921-400, Rio de Janeiro, Brazil.\\
$^2$ Instituto Federal de Educa\c c\~ao, RJ, 27230-100, Volta Redonda, Brazil.\\
$^3$ Universidade do Estado do Rio de Janeiro UERJ, RJ, 20550-013, Rio de Janeiro, Brazil.}
\date{March, 29, 2014}
\begin{abstract}

The solar quiet  daily variation ({\bf Sq}) was investigated with respect to the longitudinal and seasonal variations at the Brazilian geomagnetic ground stations of Tatuoca ({\bf TTB}), Vassouras ({\bf VSS }) and  S\~ao Martinho da Serra ({\bf SMS}). The data utilized was collected  during the time interval from January to May 2010.
We employed the continuous wavelet transforms and cross wavelet coherence to study the spectral content and correlations of the time series in the time-frequency domain. We identified possible phenomena of regional or global scope that could be affecting the magnetic response measured at {\bf TTB}, {\bf VSS} and {\bf SMS} observatories. As a result some signatures of planetary waves and semidiurnal tides in the {\bf Sq} variability were obtained.

\end{abstract}

\vspace{1cm} $\dag $ {\bf phone:}(55)(21)35049142, {\bf fax:}(55)(21)25807081.

\maketitle

\section{Introduction}

It is well know that the normal solar activity alters the currents system  in the ionosphere. It produces two magnetic effects which are prominent in the equatorial and mid-latitude regions and stronger at local noontime.  The first is called  solar quiet daily variation ({\bf Sq})  and the second is the equatorial electrojet ({\bf EEJ}).
The {\bf EEJ} takes place in a narrow belt of $\pm 3^\circ$ centered over the dip magnetic equator and about $105\, km$ high, which , flows eastward during the local daytime. Its effects at the Earth's surface range up to about $100\, nT$. The {\bf Sq} current, on the other hand, has its centres near $30^\circ$ magnetic (dip) latitude  in both hemispheres, flowing counterclockwise in the northern and clockwise in the southern hemisphere and its  magnetic effects range up to about $40\,nT$ near the Earth's surface \cite{Forbes, Rastogi, Onwumechili, Price}.
The {\bf EEJ} and the {\bf Sq} Current System interact with another atmospheric, magnetosphere and solar phenomena, e.g. planetary waves, magnetic storms and solar flares, which characterize its hemispherical asymmetries and its local and seasonal responses \cite{ALi & Pancheva_Book}.
Numerous investigations on these interaction mechanisms have been done analyzing geomagnetic data obtained from ground stations as well as from satellite magnetometers. Despite of the amount of works carried out, this theme remains the subject of active research, e.g., \cite{Padatella, Sahai}. The reasons of this interest are the requirements of more precise knowledges on the local variation of the geomagnetic field for the development of more accurate mathematical models and some still unclear aspects on the interaction mechanisms between different ionospheric phenomena.

There are former studies of the Daily Variation and the Equatorial Electrojet in the Brazilian zone using magnetic data from ground stations. These studies have been carried out basically  processing  magnetic data collected in campaigns during the International Equatorial Electrojet Year ({\bf IEEY}, $Sep.\, 1991\, –\, Mar.\, 1993$ ). These researches have covered mainly the dip equator zone and have been focused in the study of the {\bf EEJ} \cite{Rastogi_2010, Rastogi_2009, Rastogi_2007}.

In this work it is investigated the day to day variability of the solar quiet day variation in the Brazilian zone. We used data from the geomagnetic stations of Tatuoca ({\bf TTB}), Vassouras ({\bf VSS}) and S\~ao Martinho da Serra ({\bf SMS}), which cover from the dip equator to the mid-latitudes zone. The processed data covers the time interval of the first five months of the year 2010.
The aim of this research is to explore the potentiality of the present high resolution magnetic data generated in on ground magnetic observatories localized in the Brazilian sector to investigate local characteristics of these ionospheric phenomena and also to explore the advantages that, for this type of investigation, the spectral analysis in the time-frequency domain offers.

\section{Data set}

\begin{figure}[[htbp]
\includegraphics[height=6.2cm, width=8.0cm]{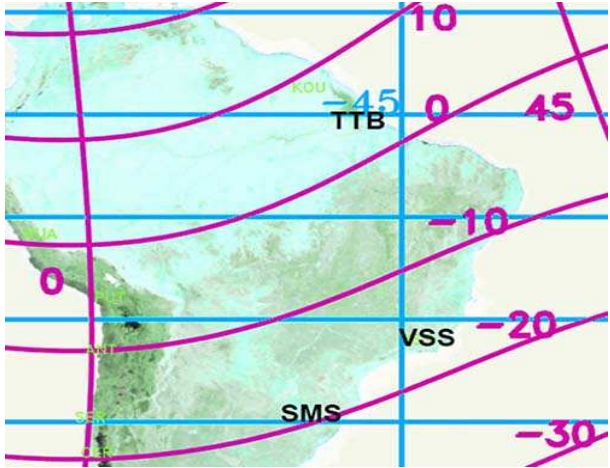}
\caption{\label{Geomagnetic_Coordinates_by_SuperMag} Map showing the locations of the relevant Brazilian observatories. Labels indicate the IAGA code of observatories. Geographic co-ordinates are: S. Martinho da Serra (-29,538, -53,855), Vassouras (-22,404, -43,663) and Tatuoca (-1,2001, -48,506).\\ Figure edited from $http://supermag.uib.no/info/img/SuperMAG_Earth_GEO.png$.}
\end{figure}

We used the data collected during the first five months of  2010 at the Brazilian Geomagnetic Observatories of Tatuoca ({\bf TTB}), Vassouras ({\bf VSS }) and  S\~ao Martinho da Serra ({\bf SMS}). These observatories comprise almost all the Brazilian zone from mid-latitudes to the dip-equator (see Figure \ref{Geomagnetic_Coordinates_by_SuperMag}).
The data consists of a record of $Hx$, $Hy$  and $Hz$ components  of the geomagnetic field with a resolution of one second, which was processed  and  smoothed using a moving average filter reducing the resolution to one hour, which is more appropriated for daily variation studies.

We have calculated the monthly means of the daily variation taking into account the international quietest days of the month, which are deduced from the planetary three-hour index Kp.  The quietest days of the first ten months of the year 2010 are listed in Table \ref{Int. Quietest Days}. These data were obtained from the International Service of Geomagnetic Indices of the International Association of Geomagnetism and Aeronomy.
\begin{table}[ht]
\caption{International Quietest Days ($ Q_1 - Q_{10} $) for the year 2010. Data from GeoForschungsZentrum Potsdam, Adolf-Schmidt-Observatorium f\"ur Geomagnetismus, Germany}
\begin{center}
\begin{tabular}{|c|c|c|c|c|c|c|c|c|c|c|c|c|c|c|c|c|} \hline
\multicolumn{12}{|c|}{~~~~~~~~~~Quietest Days} \\
\hline
Month\,\,\,\,\, Year & $Q_1$ & $Q_2$ & $Q_3$ & $Q_4$ & $Q_5$ & ~~~~~~~~~~ & $Q_6$ & $Q_7$ & $Q_8$ & $Q_9$ & $Q_10$ \\
 \hline
 Jan. ~ 2010~~~~~ & 17 &  7 &  9 &  2 &  8   & ~~~~~~~~~~ &  27 &  6 & 16 & 19  & 1  \\
 Feb. ~ 2010~~~~~ &  20 & 21 & 27 &  5 & 28  &~~~~~~~~~~ &  9 & 26 & 10 & 23 &  7 \\
 Mar. ~ 2010~~~~~ & 22 & 23 &  21  & 9  & 8  &~~~~~~~~~~ & 13 & 15 &  5 & 19 & 16 \\
 Apr. ~ 2010~~~~~ & 26 & 10 & 18 & 25 & 30   &~~~~~~~~~~  & 16 & 28 & 13 & 17 &  20 \\
 May  ~ 2010~~~~~ & 23 & 24 & 27 &  9 & 13   &~~~~~~~~~~  & 15  & 1 & 14 & 22 & 16 \\
 Jun. ~ 2010~~~~~ & 12 & 20 &  8 & 19 &  9   &~~~~~~~~~~  & 23 & 14 & 11 & 22 &  7 \\
 Jul. ~ 2010~~~~~ & 10 & 17 & 18  & 7 & 13  &~~~~~~~~~~ &  6 & 19 &  8  & 5 & 16\\
 Aug. ~ 2010~~~~~ & 30 & 22 & 21 & 29 & 14   &~~~~~~~~~~  & 20 & 31 & 13 & 19 & 7 \\
 Sep. ~ 2010~~~~~ & 11 & 12 & 30 &  4 & 22  &~~~~~~~~~~ & 10 &  3 & 13 & 19 & 20  \\
 Oct. ~ 2010~~~~~ &  2 & 1 & 4 &  1 &  3   & ~~~~~~~~~~ & 30 & 21 &  7 & 28 & 10 \\
\hline
\end{tabular}
\end{center}

\label{Int. Quietest Days}
\end{table}

\section{The Daily Variation}

Here we are interested in the comparison of the local and seasonal response of the amplitude and phase of the daily variation ({\bf Sq}) among the three geomagnetic observatories chosen for this study and to detect eventual occurrences of abnormal phases, i.e.,  when the maximum of {\bf Sq} is reached out of the time interval [8:30-13:30]. For this purpose in this section we calculated the monthly means of the horizontal and vertical components for the five quietest days of each month.
Former studies reveal the occurrences of  abnormal variations of the  phase and amplitude of the day to day variability of the {\bf Sq} . In mid-latitudes the occurrence of these events has been attributed to northward and southward perturbation of the horizontal component of the geomagnetic field of large geographical extend and possible magnetospheric origin \cite{Butcher}. On the other hand for the dip latitude this phenomenon has been attributed to the occurrence of partial equatorial counter-electrojet currents due to changes in local  ionosphere dynamos. Other local factors could contribute to the regional variability of the solar quiet day variation, e.g., induction by the onshore horizontal component at the ocean-continent interface, oceans induction and conductivity anomalies \cite{Lilley}.

The calculated monthly means of the quietest day variation are plotted in Figure \ref{Monthly Average in TTB, VSS, SMS}. The data covers  the period of Jan-May 2010.  The left column shows the values of the horizontal component and the right column the vertical component. The rows indicate, from top to bottom, the ground stations of {\bf TTB}, {\bf SMS} and {\bf VSS}. The curves show visible seasonal and latitudinal dependencies  for the amplitude and phase of {\bf Sq}. Here the phase is understand as the time at which the maximum amplitude of {\bf Sq} is reached.
\begin{figure}[[htbp]
\includegraphics[height=9.3cm, width=15.5cm]{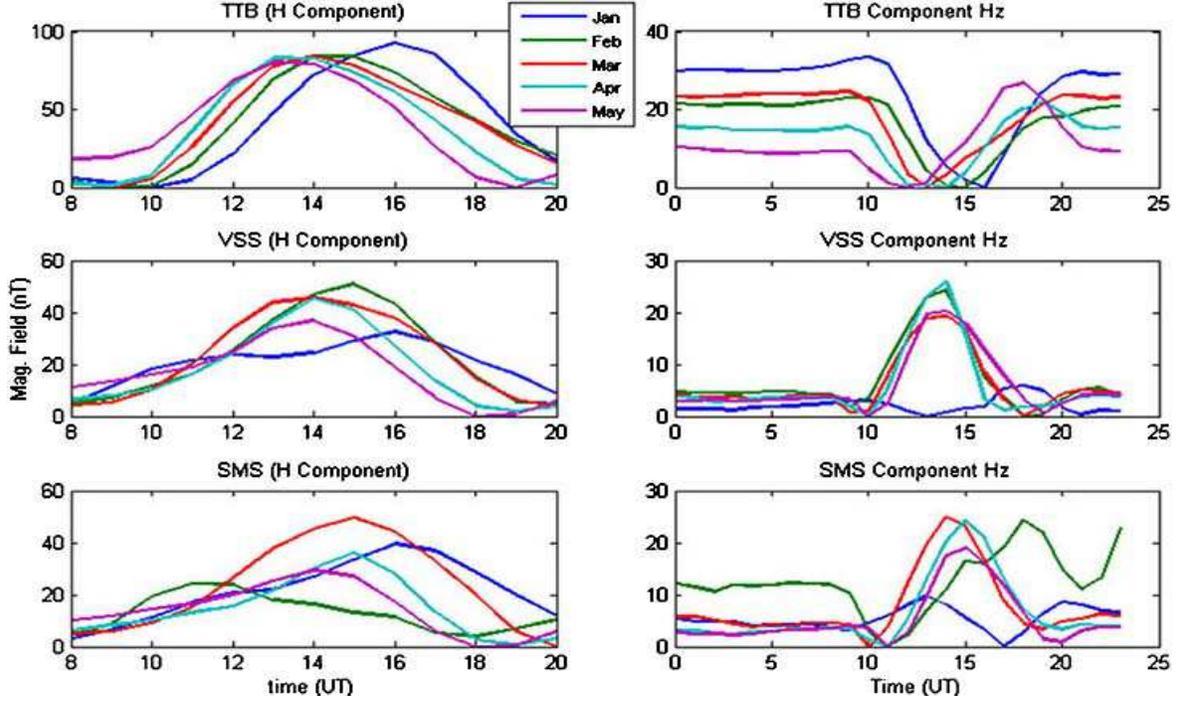}
\caption{\label{Monthly Average in TTB, VSS, SMS} Monthly average of {\bf Sq} at Tatuoca, Vassouras and S. Martinho da Serra.}
\end{figure}
In Figure \ref{Monthly Average in TTB, VSS, SMS} it can be observed an enhancement of the {\bf Sq} signal at the Tatuoca station. This is a well known effect due to the equatorial electrojet phenomenon. It can be noted also that the horizontal component at {\bf TTB} station shows its earliest phase and its minima amplitude in May and undergoes the maxima amplitude variability and latest phase in January. However, for {\bf VSS} and {\bf SMS} stations, the latest phase follows solely this behavior. At the same time it can be observed that the spread of the phase in {\bf VSS} is larger than in {\bf TTB} and smaller than in the {\bf SMS} station. That is, the phase spread increases in the meridional direction. The vertical component in {\bf TTB} follows a seasonal behavior similar to the horizontal one, but for {\bf VSS} and {\bf SMS} this component exhibits an abnormal lower amplitude variation in January.
Former studies \cite{Butcher, Hitchman} have asserted that mid-latitude stations tend to have maximum {\bf Sq} amplitudes (and advanced phase) at summer solstice, while  at equatorial stations this tendency takes place at equinoctial times. In our case the results for {\bf VSS} and {\bf SMS} reveal that this maximum occurs for summer equinoxes and for the {\bf TTB} observatory it is reached at the summer solstice.

Local asymmetries and atypical seasonal behavior of these ionospheric phenomena are expected to occur due to non-migrating tides and asymmetric tidal winds driven by the hemispheric asymmetries of the conductivity due to variations in the insolation particulary during the solstice season \cite{Padatella}.

\begin{figure}[[htbp]
\includegraphics[height=9.3cm, width=13.5cm]{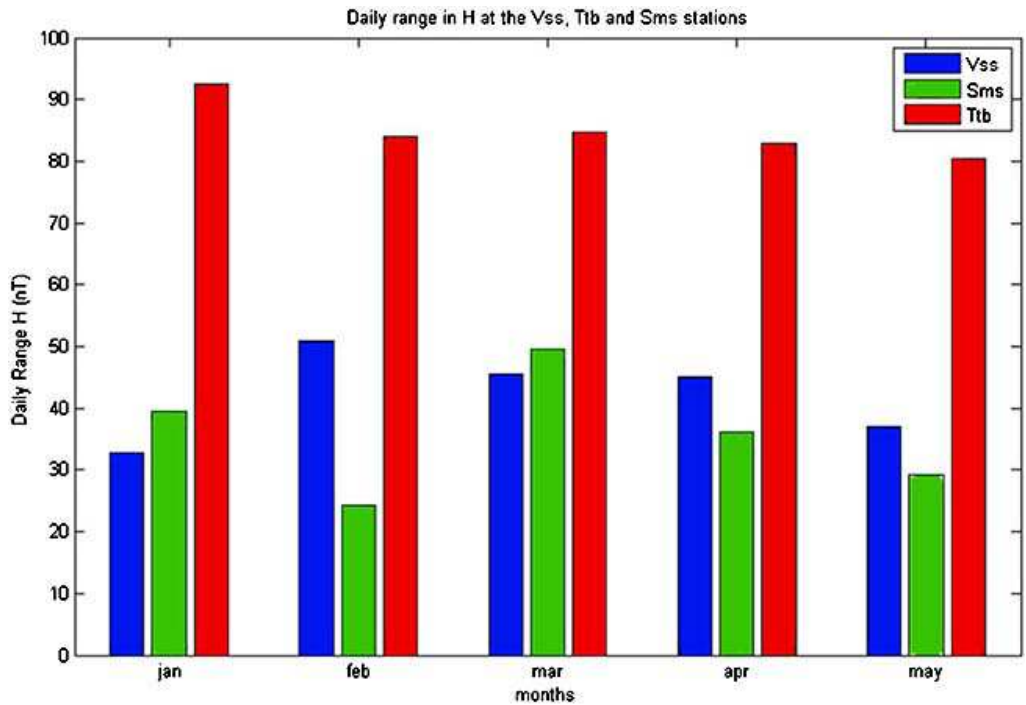}
\caption{\label{BarPlot Daily Range at VSS SS TTB jan-may 2010}  Comparison of the monthly means of the Daily Range at the stations of {\bf TTB}, {\bf VSS} and {\bf SMS} from January to May of 2010.}
\end{figure}

The bar plot of Figure \ref{BarPlot Daily Range at VSS SS TTB jan-may 2010} shows the daily range of {\bf Sq} signal at the three stations for the first five months of 2010. In Figure \ref{BarPlot Daily Range at VSS SS TTB jan-may 2010} it can be observed a decreasing tendency of the {\bf Sq} range at Tatuoca observatory, exhibiting its higher value at January, while for the other two stations  it shows a bell shape with its maximum value in March. As can be expected this result confirms the same seasonal behavior shown in Figure \ref{Monthly Average in TTB, VSS, SMS}.
Comparing the values of the daily range in the Vasouras and S. Martinho da Serra it can be appreciated a monthly alternation of its maximum value. This seems to arise by coincidence because this parameter is sensitive to the Lloyds Seasons, i.e., to the equinoctial and solstice months, rather than to isolate months.

\begin{figure}[[htbp]
\includegraphics[height=9.3cm, width=9.5cm]{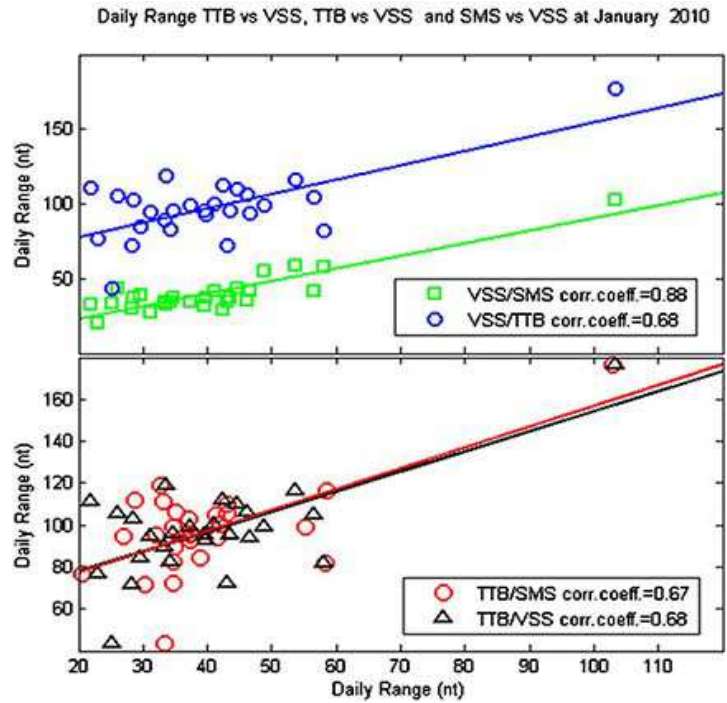}
\caption{\label{Mass_Plot_Daily_Range_TTBvsVSS_TTBvsSMS_SMSvsVSSJan2010} Comparison of the values of the Daily Range among the three stations from Jan-May 2010.}
\end{figure}

Figure \ref{Mass_Plot_Daily_Range_TTBvsVSS_TTBvsSMS_SMSvsVSSJan2010} shows two scatter plots for different stations pairs. The top panel shows a comparison of the daily variation at {\bf VSS} station with respect to the other two. Here it can be seen that the better correlation is achieved for the station pair {\bf VSS/SMS} with a correlation coefficient ({\bf CF}) of 0.88, which falls for the station pair {\bf TTB/VSS} to $CF = 0.68$. This is an expected result due to their geographical locations.
The bottom panel displays the comparison of this parameter at {\bf TTB} related to {\bf VSS} and {\bf SMS} stations which are situated at a large distance with respect to the first one. Here it can be observed that in both cases the correlation is lower.
This suggests that the seasonal response of the daily range at Tatuoca station, which is clearly different with respect to the other two, arises principally due to the additional contributions of the equatorial electrojet, which are absent at the other two stations.
On the other hand the correlation of the daily range between Vassouras and S. Martinho da Serra is 0.88, which says that the bell shape of the seasonal response, showed by both stations, is due to the influence of the daily variation current and other local factors.

\section{Wavelet Spectral Analysis}

Wavelet spectral analysis allows us a quantitative monitoring of the signal evolution by decomposing a time-series into time-frequency space. In this way  some problems associated to the Fourier method can be avoided, determining both the dominant modes of variability and how those modes vary in time. Wavelets are useful especially for signals that are non-stationary, have short-lived transient components, have features at different scales or have singularities \cite{Wavelet&Applications}.

For the computation of the continuous wavelet transform ({\bf WT}) we used a free script initially developed for oceanographic research \cite{torrence, Torrence_Script}. We use the Morlet wavelet to perform the spectral analysis. It is the better choice, when one is interested to look for special prints from other oscillatory phenomena, because this wavelet shows an acceptable localization in both time and frequency.

\begin{figure}[[htbp]
\includegraphics[height=9.3cm, width=15.5cm]{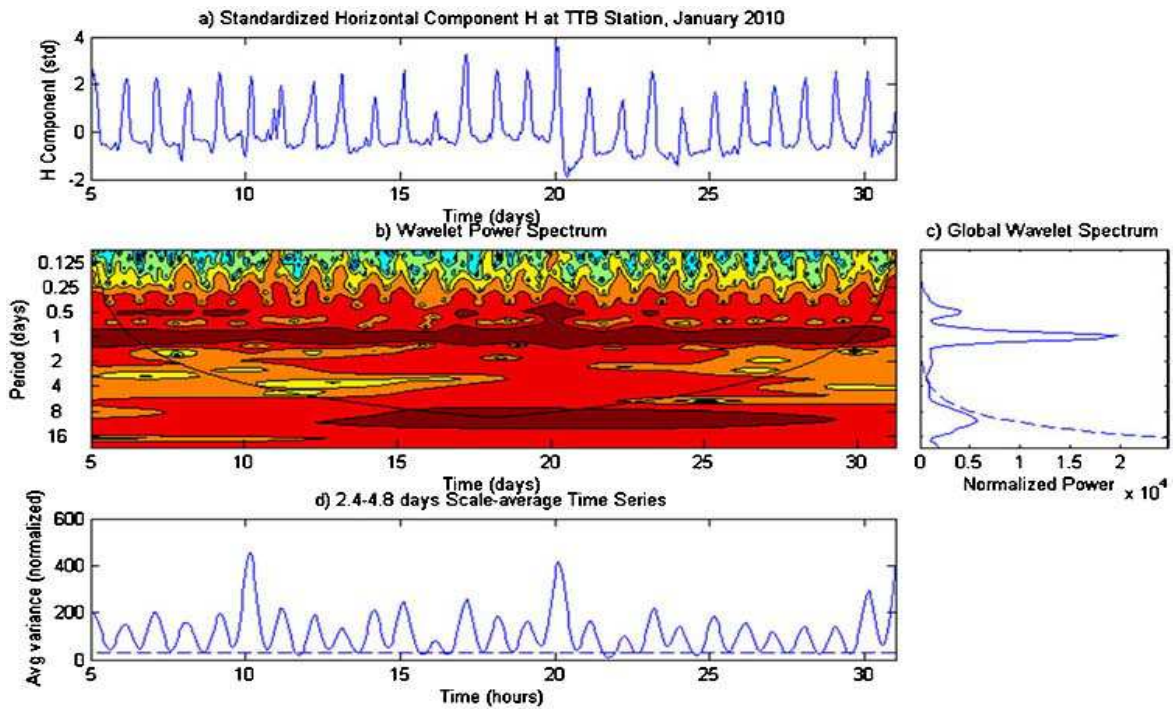}
\caption{\label{CWT TTB} Wavelet Power Spectrum of the Magnetic Horizontal Component at Tatuoca Station for January 2010.}
\end{figure}

Figure \ref{CWT TTB} shows the calculated wavelet power spectrum of the horizontal component at Tatuoca. In this case the time series corresponds to the values measured during the month of January 2010.
Here it can be observed a well defined peak corresponding to the daily variation during the whole month. In addition, it can be noted other two components corresponding to the semidiurnal period and to the 6-8 days period respectively. The semidiurnal signal arises in the form of wave trains around the 6th and 20th days. The 6-8 days period appears as an isolated peak, in the limit of the influence cone, also around the day 20. This day corresponds to one of the most disturbed days of the  month.

\begin{figure}[[htbp]
\includegraphics[height=9.3cm, width=15.5cm]{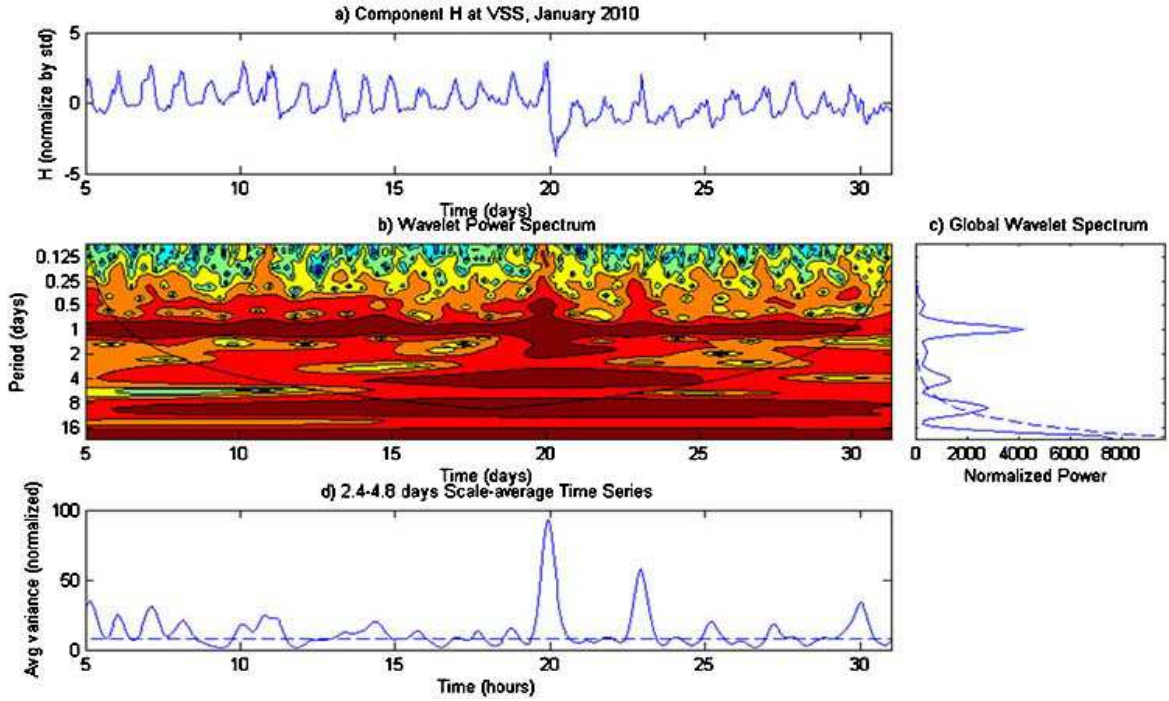}
\caption{\label{CWT VSS} Wavelet Power Spectrum of the Magnetic Horizontal Component at Vassouras Station for January 2010.}
\end{figure}

\begin{figure}[[htbp]
\includegraphics[height=9.3cm, width=15.5cm]{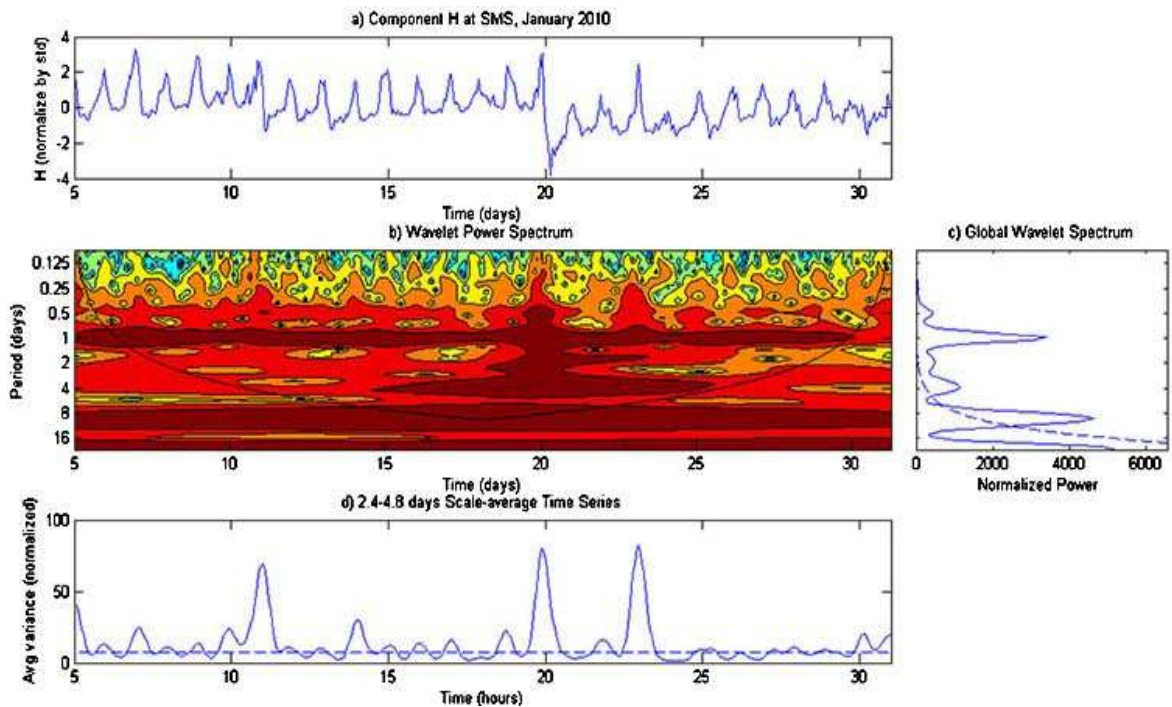}
\caption{\label{CWT SMS} Wavelet Power Spectrum of the Magnetic Horizontal Component at  S. Martinho da Serra Station for January 2010.}
\end{figure}

In figures \ref{CWT TTB}, \ref{CWT VSS} and \ref{CWT SMS} the wavelet spectral analysis for the signal at Tatuoca,  Vassouras and S. Martinho da Serra stations for January 2010 are respectively presented. In all three stations, as should be expected, the component of the signal corresponding to the daily variation during all the month appears. The semidiurnal component at {\bf TTB} station is visible during the first days of the month and around the 20th day, however at the others two stations the signal emerges only associated to the most disturbed day of the month. At the stations of {\bf VSS} and {\bf SMS} (Figures \ref{CWT VSS} and \ref{CWT SMS}) components of the signal corresponding to the 2 days, 4 days and 6-8 days periods are visible. These oscillations arise also associated to the disturbed days. For the {\bf TTB} station (Figure \ref{CWT TTB}) only components for the 6-8 days periods are present.

The Cross Wavelet Transform ({\bf XWT}) and the wavelet coherence ({\bf WTC}) have been used to analyze geophysical time series. Given two times series $x_n$ and $y_n$ and their respective continuous wavelet transforms $Wx$ and $Wy$ the {\bf XWT} is defined as $W_{xy}\,=\,W_{x}\,W_{y}^*$. As the {\bf XWT} is a complex magnitude it is useful to define the cross wavelet power as $|W_{xy}|^2$ and the phase difference as $arg(W_{xy})$, which can describe the common power and relative phase between $x_n$ and $y_n$ in time-frequency space. See \cite{Grinsted} for details.

The concept of coherence between two time series appeared initially extending the definition of the cross correlation using notions of grade of coherence in classical optics \cite{Wiener}. The Wavelet Coherence  is a natural extension to the time-frequency domain of this former definition . It tells us how much coherent the Cross Wavelet Spectrum is, allowing us to seek confirmation of causal links between both time series.

We have analyzed two pairs of time series in order to test possible relationships between them. The first pair  formed by the signals from {\bf VSS} and {\bf SMS} geomagnetic Stations and the second one formed by the signals from {\bf VSS} and {\bf TTB} stations. The power wavelet transform of each separated time series was examined above (see figures \ref{CWT TTB}, \ref{CWT VSS} and \ref{CWT SMS}) and there were detected some common characteristics that we will checkout with this method.

\begin{figure}[[htbp]
\includegraphics[height=9.3cm, width=15.5cm]{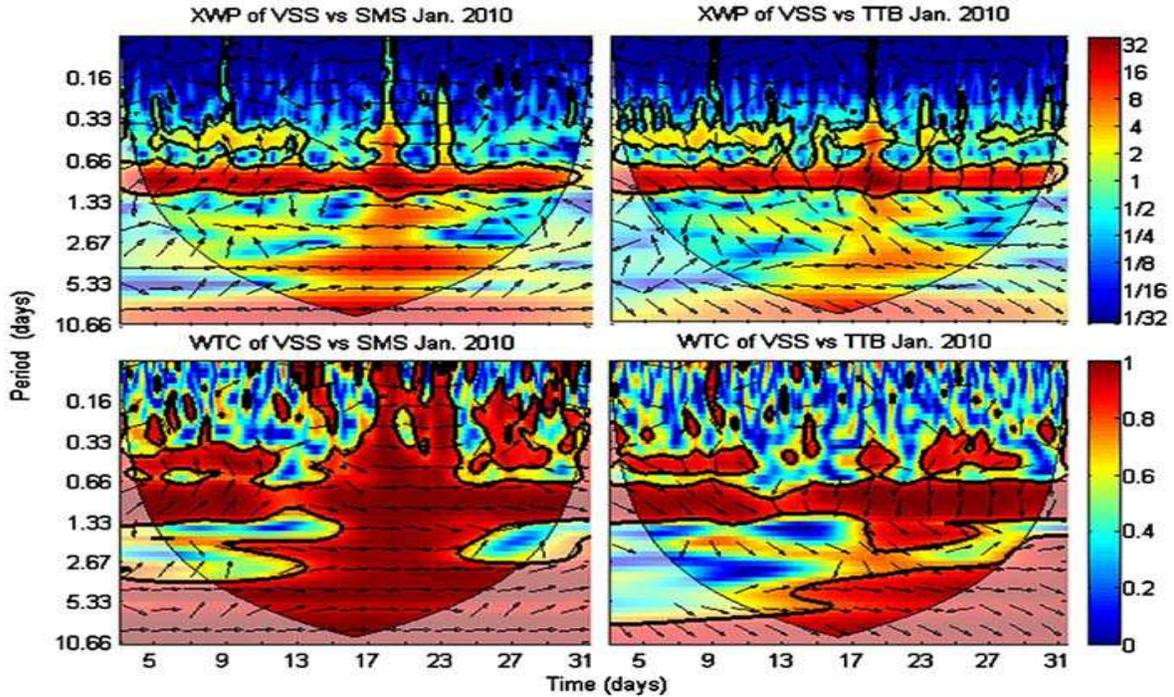}
\caption{\label{XWT and WTC} Cross Wavelet Power Spectrum (upper panel) and  Wavelet Coherence for the pairs: {\bf VSS} vs. {\bf TTB} (right) and {\bf VSS} vs. {\bf SMS} (left), for Janurary 2010. The arrows indicate the relative phase. Is indicated also the cone of influence, with the thick contour designating the $ 5\% $ significance level. }
\end{figure}

Figure \ref{XWT and WTC} shows the standardized cross power spectrum ({\bf XWP}) and the Wavelet Coherence of signals pairs for January 2010. It is also indicated the $5\%$ confidence level by a thick contour and the influence cone.
In Figure \ref{XWT and WTC} it can be noted, as expected, more significative zones of common power for the pair {\bf VSS-SMS} than for the {\bf VSS-TTB} one. The {\bf XWP} of the pairs {\bf VSS-SMS} and {\bf VSS-TTB} are shown in the upper-right and upper-left panels respectively. In these two plots it can be appreciated that both pairs exhibit a significative common power, during all the month, for the period corresponding  to the daily variation. For the semidiurnal period it is also observed a significant common cross power, but only for some time intervals.
For the {\bf VSS-SMS} pair it appears for the 5\textsuperscript{th} to 12\textsuperscript{th} days and around the 17\textsuperscript{th} to 21\textsuperscript{th} days. For the {\bf VSS-TTB} pair this zone is extended to the last week of the month. For the {\bf XWP} of the {\bf VSS-SMS} pair, besides these significant zones of common power, other zones of common power centered around periods of $2$, $4$ and $8$ days also appear. For the {\bf VSS-TTB} pair the common {\bf XWP} in these zones appears weaker and only survives the zone around the period of $8$ days. However, as these zones are below the confidence level, we can only speculate about causal links between time series for those periods.

Lower panels of Figure \ref{XWT and WTC} display the wavelet transform coherence ({\bf WTC}) of the signal pairs {\bf VSS-SMS} (left) and {\bf VSS-TTB} (right). In both panels it can be seen that the zone of higher coherence, above the confidence level, covers a greater area comparing with the corresponding area calculated for the Cross Wavelet Power Spectra ({\bf XWP}). For example, the zone around the 2 and 8 days periods corresponding to the time interval from the 19\textsuperscript{th} to 25\textsuperscript{th} days now appears inside the confidence zone for both pairs. However for the zone around the 4 days period it appears included for the {\bf VSS-SMS} pair, but not for the {\bf VSS-TTB} one. It also appears more evident the coherence for the semidiurnal period that now includes also the last week for the {\bf VSS-SMS} pair.

In Figure \ref{XWT and WTC} an arrows mesh indicates the relative phase between time series. When the arrows point to the right indicate that both time series are in-phase and the first series leads the second one. Arrows pointing to left means that the series are in anti-phase and the first series lags the second one.
For the {\bf VSS-SMS} pair it can be observed that time series are in phase for all the periods, except for the time interval from the 5\textsuperscript{th} to the 13\textsuperscript{th} days, where in the zone of the semidiurnal period the time series at {\bf VSS} station lags the corresponding one at {\bf SMS} and for the zone of diurnal period, where the series at {\bf VSS} leads the series at {\bf SMS}. For the {\bf VSS-TTB} pair it can be observed, for the diurnal period, that the time series at {\bf VSS} station leads the corresponding one at {\bf TTB} station during the first 12 days. Around the 15\textsuperscript{th} day, the series at {\bf VSS} starts to follows the series at {\bf TTB} and remains, for the rest of the month, following the series at {\bf TTB} with a relative phase of $90^\circ$. For the zone comprising the periods from $4$ to $8$ days and the time interval from the 15\textsuperscript{th} to 20\textsuperscript{th} days, the series at {\bf VSS}  seems to lead the series at {\bf TTB} with a small relative phase.

\section{Discussion}

Employing the cross wavelet and wavelet coherence we detected common oscillations, with periods around the $2\,$, $4\,$ and $8\,$ days, in the horizontal magnetic component for the three magnetic observatories. Taking into account the geographic distances between these stations and its periods it can be inferred that such oscillations are due to planetary waves.
In Figure \ref{XWT and WTC} it can be observed that the oscillations around the $2\, d$, $4\, d$ and $8\, d $ periods appeared associated to the days 10\textsuperscript{th} to 21\textsuperscript{st}, which were the more perturbed days of the analyzed time interval. Oscillations of such periods, connected also to disturbed days, have been observed from meteor radar measurements of mesosphere and ionosphere of the Brazilian equatorial zone \cite{Takahashi} and in the northwestern Pacific zone \cite{Yuji}.

In Figure \ref{XWT and WTC} the wavelet coherence spectrum shows Quasi-2d oscillations ({\bf QTDO}), within the confidence zone, for the {\bf VSS-SMS} pair. This result concords with the fact that in the Southern Hemisphere the occurrence of {\bf QTD} stratospheric planetary waves has a maximum around January \cite{Plumb}.
However the {\bf VSS-TTB} pair does not exhibit such oscillation, because the wavelet power spectrum of the signal at the {\bf TTB} station (see Figure \ref{CWT TTB}) does not hold significative spectral content at these periods. This is an unusual fact since {\bf QTDO} are more frequent at the dip equator than at mid-Latitudes. {\bf QTDO}  have been reported in the equatorial electrojet ({\bf EEJ}), with a peak at late solstice months (i.e., during January, February, July and August) \cite{Gurubaran}. The fact that we have not found them at {\bf TTB} could be explained by the fact that such oscillations in the {\bf EEJ} are not always linked with planetary waves.

\section{Summary and Conclusions}

The daily variation was studied during the first five months of 2010 using the data form three geomagnetic observatories located in the Brazilian zone. Well known results were reproduced, e.g., the enhancement of {\bf Sq} and the seasonal behavior of the daily range at the equatorial electrojet zone. However an anomalous oscillatory seasonal behavior was obtained for the daily range for the stations of {\bf VSS} and {\bf SMS}. The seasonal response obtained for the phase of  maximum {\bf Sq} amplitude also differs from former results.

Employing the wavelet analysis zones of significant spectral power for the periods of $1/2\,d$ and $1\,d$ which are related with the diurnal and semidiurnal tides were detected. There were also obtained contributions  in the {\bf Sq} response from oscillation around the $2\,d$, $4\,d$ and $8\,d$ periods, which can be interpreted as signatures of planetary waves on the {\bf Sq} response. It should be noted that the spectral analysis was carried out using a relatively small data set. A future research must involve others geomagnetic observatories and larger time series in order to improve the characterization of the seasonal behavior of such oscillations in the geomagnetic data.

\begin{acknowledgments}

We acknowledge Potsdam Geophysical observatory GFZ for the {\em Solar Quiet Days} databases available at GFZ website. Wavelet software was provided by C. Torrence and G. Compo, and is available at URL: http://paos.colorado.edu/research/wavelets/. "Crosswavelet and wavelet coherence software were provided by  A. Grinsted." (C) Aslak Grinsted 2002-2004 available at:  http://www.pol.ac.uk/home/research/waveletcoherence/.   We acknowledge SuperMAG initiative for the maps available at website $http://supermag.uib.no/info/img/SuperMAG_Earth_GEO.png.$

This material is based upon work supported by the {\bf DTI-PCI} fellowship $N.380.739/09-7$ of Brazilian Science Funding Agency. A. R. R. Papa wishes to acknowledge {\bf CNPq} (Brazilian Science Foundation) for a productivity fellowship and FAPERJ (Rio de Janeiro State Science Foundation) for partial support.
\end{acknowledgments}


\end{document}